\documentclass[runningheads]{llncs}
\usepackage[T1]{fontenc}
\usepackage{graphicx}
\usepackage{amsmath,amsfonts}
\usepackage{algorithmic}
\usepackage{algorithm}
\usepackage{array}
\usepackage{textcomp}
\usepackage{stfloats}
\usepackage{url}
\usepackage{verbatim}
\usepackage{graphicx}
\usepackage{cite}
\usepackage{multirow} 
\usepackage{float}
\usepackage{colortbl}
\graphicspath{ {image/} }
\usepackage{vcell}
\usepackage{adjustbox}
\usepackage{url}
\usepackage{makecell}
\usepackage{multirow}
\usepackage{rotating}
\usepackage{threeparttable} 
\usepackage{amsmath}  
\usepackage{amsfonts,amssymb}

\usepackage{booktabs}
\usepackage{arydshln}
\usepackage[misc]{ifsym}
\usepackage{xcolor}
\usepackage{cite}
\usepackage[normalem]{ulem}
\usepackage{multirow}

\usepackage{xspace}
\newcommand{\etal}{\emph{et al.}\xspace}

\begin{document}
\title{Unveiling Latent Information in Transaction Hashes: Hypergraph Learning for Ethereum Ponzi Scheme Detection}
\titlerunning{Unveiling Latent Information in Transaction Hashes}

%
\author{
Junhao Wu\inst{1} \and
Yixin Yang\inst{2} \and
Chengxiang Jin\inst{1,3} \and 
Silu Mu\inst{4} \and
Xiaolei Qian\inst{4} \and \\
Jiajun Zhou\inst{1,3} \textsuperscript{(\Letter)} \and 
Shanqing Yu\inst{1,3} \and
Qi Xuan\inst{1,3} 
}

\authorrunning{Wu \etal}
%
\institute{
Zhejiang University of Technology, Hangzhou 310023, China  
\and
CNCERT/CC, Beijing 100029, China
\and
Binjiang Institute of Artificial Intelligence, ZJUT, Hangzhou 310056, China
\and
Servyou Information Technology Co., Ltd, Hangzhou 310053, China
\email{jjzhou@zjut.edu.cn}}
%
\maketitle              
\begin{abstract}

With the widespread adoption of Ethereum, financial frauds such as Ponzi schemes have become increasingly rampant in the blockchain ecosystem, posing significant threats to the security of account assets. Existing Ethereum fraud detection methods typically model account transactions as graphs, but this approach primarily focuses on binary transactional relationships between accounts, failing to adequately capture the complex multi-party interaction patterns inherent in Ethereum. To address this, we propose a hypergraph modeling method for the Ponzi scheme detection method in Ethereum, called HyperDet. Specifically, we treat transaction hashes as hyperedges that connect all the relevant accounts involved in a transaction. Additionally, we design a two-step hypergraph sampling strategy to significantly reduce computational complexity.  
Furthermore, we introduce a dual-channel detection module, including the hypergraph detection channel and the hyper-homo graph detection channel, to be compatible with existing detection methods. 
Experimental results show that, compared to traditional homogeneous graph-based methods, the hyper-homo graph detection channel achieves significant performance improvements, demonstrating the superiority of hypergraph in Ponzi scheme detection.
This research offers innovations for modeling complex relationships in blockchain data.

\end{abstract}

\section{Introduction}
As the leading smart contract platform, Ethereum~\cite{tikhomirov2018ethereum} has fueled the rapid development of emerging fields such as decentralized finance~\cite{sriman2022decentralized} through its decentralization, transparency, and programmability. 
The automated execution feature of smart contracts~\cite{wang2021ethereum} allows users to perform complex financial operations, including lending, trading, and investing, without third-party intermediaries. 
However, the widespread application of Ethereum within the financial sector has also created opportunities for fraudulent activities~\cite{chen2020survey} due to its anonymity and decentralized nature. 
Ponzi schemes~\cite{wilkins2012understanding}, in particular, lure investors with promises of high returns, ultimately resulting in significant financial losses for users and posing a serious threat to the development of the Ethereum ecosystem.

Existing fraud detection methods~\cite{pourhabibi2020fraud,jin2022dual} rely on homogeneous graph modeling and make some progress. This modeling approach is intuitive but has significant limitations. 
First, homogeneous graphs focus only on binary transaction relationships between accounts. This modeling approach does not sufficiently capture the inherent complexity of multi-party interaction patterns in Ethereum, such as those involving multiple account interactions during contract calls~\cite{chen2020traveling}. 
This limitation prevents traditional methods from fully modeling the complete interaction features between accounts, thereby reducing fraud detection performance. Moreover, traditional Graph Neural Networks (GNNs)~\cite{wu2020comprehensive} primarily depend on information propagation between nodes and their immediate neighbors, making it challenging to model higher-order dependencies between accounts, such as indirect transaction chains or fund flow patterns formed through multiple intermediary accounts~\cite{chen2020understanding}. 
These higher-order interactions are often critical features of fraudulent activities like Ponzi schemes. 

To address the above issues, this paper proposes a hypergraph~\cite{antelmi2023survey} modeling method for Ponzi scheme detection in Ethereum, called \textbf{HyperDet}. In particular, we focus on Ethereum transaction hashes, which contain multi-party interaction information and can naturally be constructed as hypergraphs. 
In addition, this paper designs an efficient two-step hypergraph sampling strategy consisting of hyperedge filtering and node refinement. 
The former samples the hyperedges around the target node, while the latter samples the nodes within the hyperedges. These sampling methods significantly reduce the computational complexity, enabling the model to handle large transactional data. 
Finally, we introduce a dual-channel detection module, called the hypergraph detection channel and the hyper-homo graph detection channel. 
The former uses hypergraph neural networks (HGNNs) to capture higher-order information in the hypergraph to learn complex behavioral patterns, while the latter captures the information by converting the hypergraph into a hyper-homo graph and following a GNN algorithm. 
Experimental results demonstrate the efficacy of the proposed method, outperforming homogeneous graph-based approaches across multiple evaluation metrics. This provides a novel solution for Ethereum fraud detection. The primary contributions of this paper can be enumerated as follows:

\begin{itemize}
    \item[$\bullet$] We propose hypergraph modeling based on transaction hashes that effectively capture multi-party interaction patterns in the Ethereum network. 
    To the best of our knowledge, hypergraphs are not extensively adopted in Ethereum. Our work explores the practical and potential benefits of hypergraphs for fraud detection in Ethereum.
    
    \item[$\bullet$]We design a two-step sampling method to address the challenges of the vast Ethereum transaction network.
    The sampling approach significantly reduces the computational complexity while preserving the integrity.
    
    \item[$\bullet$] We design a dual-channel detection module and introduce a method for converting hypergraphs into hyper-homo graphs, thus being compatible with existing graph-based detection methods.

    \item[$\bullet$] Extensive experiments indicate that the hypergraph-based detection method offers higher performance in identifying Ponzi schemes, providing a strong safeguard for the security of the Ethereum ecosystem.
\end{itemize}


\section{Related Work}
\label{sec: related work}
\subsection{Fraud Detection in Ethereum}

Initial studies primarily focused on feature engineering and machine learning approaches. 
Chen et al.\cite{chen2018detecting} extracted features from smart contract bytecodes and transactions, and achieved fraud detection with machine learning algorithms. 
Zhang et al.~\cite{zhang2021detecting} proposed an enhanced LightGBM-based~\cite{ke2017lightgbm} method, which demonstrated superior detection performance for Ponzi scheme detection.
While these machine learning approaches enabled efficient detection, they were constrained by manual feature quality and weak in modeling complex interaction patterns. 
To learn the behavior patterns, Xiong et al.~\cite{xiong2024ethereum} introduced graph modeling and developed the TransWalk algorithm, which extracts multi-scale features through random walk strategies and employs downstream machine learning classifiers for detection. 
Wu et al.~\cite{wu2020phishers} extended random walk methods by incorporating transaction amounts and timestamps, simulating dynamic propagation processes in transaction networks via biased walks. 
This approach effectively captured inter-account correlation features and achieved high performance using SVM~\cite{hearst1998support}. 
Although these graph-based methods improved performance, they neglected intrinsic account characteristics. 
With the rapid advancement of GNNs, researchers began exploring their application in Ethereum fraud detection. 
Chen et al.~\cite{chen2020phishing} generated interaction subgraphs via random walks and employed GCN~\cite{kipf2017semi} to learn node features and network topology for end-to-end detection.
Tharani et al.~\cite{tharani2024unified} introduced an edge feature aggregation mechanism to capture node structural features and dynamic behavioral patterns, 
and achieving representation learning with GraphSAGE~\cite{hamilton2017inductive}.
Duan et al.~\cite{duan2022phishing} constructed initial features from the transaction flows of target nodes and employed GraphSAGE and GCN to learn network topology. They derived the graph's final representation through a pooling layer and ultimately mapped the representation to a classification space using an MLP~\cite{popescu2009multilayer} to achieve fraud detection.
Existing methods focus on local interaction pattern representation learning and have limitations in capturing higher-order relationships.
Jin et al.~\cite{jin2022heterogeneous,jin2024time} enhanced GNN-based detection capabilities  by extracting higher-order information in meta-paths, thus demonstrating the effectiveness of higher-order information. 
However, meta-paths require domain expertise and lack generalizability. 
To address these limitations, we propose a hypergraph modeling approach via transaction hashes, which enable precise modeling of higher-order behavioral patterns and offer novel solutions for modeling Ethereum interaction networks.

\subsection{Hypergraph-based Detection Methods}

Hypergraphs, as a more sophisticated modeling paradigm, break the limitation that only two nodes can be connected by edges.
Hyperedges can connect multiple nodes at the same time, thus enabling a more flexible representation of complex multi-party interaction patterns in Ethereum, such as smart contracts call.
Currently, hypergraph neural networks (HGNNs) have gained extensive applications across diverse domains including social network analysis, recommendation systems, and bioinformatics. 
Feng et al.~\cite{feng2019hypergraph} pioneered a hypergraph neural network model termed HGNN, which synergizes hypergraph modeling capabilities with neural network feature learning. They first aggregate node features into hyperedges to capture multi-party relationships, then propagate hyperedge features back to nodes for effective high-order information updating, thereby obtaining the representation for downstream tasks.
In contrast to HGNN-based approaches, Yadati et al.~\cite{yadati2019hypergcn} proposed HyperGCN, which involves transforming the hypergraph into a simple graph and combining it with spectral domain convolution. This approach follows the proven framework of GCN while preserving higher-order information and achieving a balance between computational efficiency and performance.
Dong et al.~\cite{dong2020hnhn} developed the HNHN, enhancing model expressiveness by incorporating non-activated transformations during node-hyperedge message passing. They further designed a dataset-aware normalization scheme to dynamically adjust the significance of high-cardinality hyperedges and high-degree nodes. This approach mitigates the fixed bias toward high-cardinality hyperedges or high-degree nodes inherent in conventional methods, significantly improving model adaptability.
Nevertheless, the implementation of HGNNs in the context of Ethereum remains in its nascent stages. In this paper, we pioneered a novel approach by modeling Ethereum interaction hypergraphs and leveraging HGNNs to facilitate fraud detection.

\section{Ethereum Graph Modeling}

\subsection{Ethereum Data Overview}

Accounts and interactions on the Ethereum collectively constitute the Ethereum financial ecosystem. Firstly, a transaction hash is a unique identifier for each transaction within the Ethereum network. This hash is generated by applying a cryptographic hash function to the transaction content, ensuring the uniqueness of the transaction. The transaction hash enables the retrieval of detailed transaction information, including the senders, recipients, and transaction values. 
Secondly, an Externally Owned Account (EOA) is a user-controlled account managed via a private key, primarily used for initiating transactions. 
In contrast, a Contract Account (CA) is governed by a smart contract code and is typically created by an EOA through a transaction. 
These accounts can execute predefined logical operations, such as facilitating transfers or updating states, thereby enabling complex functionalities within the Ethereum network.
A transaction is the basic unit of interaction between accounts in the Ethereum network. A transaction can be either a transfer of Ether or a call to a smart contract function. The former involves an Ether transfer between EOAs, the latter involves an EOA calling a function in a CA, which triggers the execution of a smart contract function. Under a transaction hash, there may be multiple contract calls and account transactions to fulfill complex business logic requirements.
By analyzing them, we can gain a deeper understanding of the Ethereum network and provide powerful support for fraud detection and behavior analysis.

\begin{figure}[t]
    \centering\includegraphics[width=\textwidth]{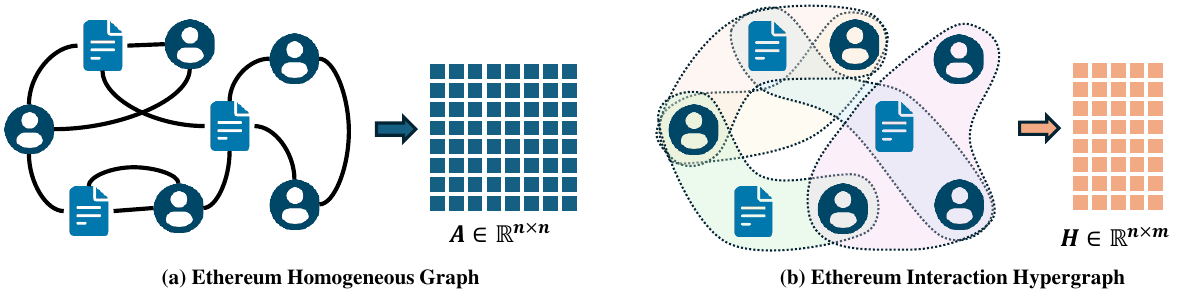}
    \caption{The Overview of Ethereum Graph Modeling.}
    \label{fig1}
\end{figure}

\subsection{Ethereum Homogeneous Graph Modeling}
In Ethereum fraud detection research, existing Ponzi scheme detection methods usually model data as a homogeneous graph, which regards accounts and their interactions as nodes and edges, respectively.
The detection method based on homogeneous graph modeling focuses on the direct interactions between account pairs, as shown in Fig.~\ref{fig1}(a). 
We express it as follows $G = (\mathcal V,\mathcal E, \boldsymbol A,\boldsymbol X,\mathcal Y )$. 
$\mathcal V = \{v_1,v_2, \ldots, v_n\}$ is a set of account nodes, irrespective of account type. 
$\mathcal E = \{e_{ij}|e_{ij}= (v_i,v_j),v_i,v_j \in \mathcal{V}\}$ is a set of interaction edges, irrespective of edge type. 
$\boldsymbol A\in \mathbb{R}^{n \times n}$ denotes the adjacency matrix. 
$\boldsymbol X = \{x_1,x_2,\ldots,x_n\}^\top \in \mathbb{R}^{n \times d}$ is referred to as the account feature matrix, where $x_i$ is the feature vector of account $v_i$ and $d$ is the dimension of the feature. $\mathcal Y = \{(v_i,y_i)|v_i \in \mathcal V_l,|\mathcal V_l|\ll n \}$ is a set of partially labeled account identities, where $\mathcal V_l$ is the set of labeled accounts, and $y_i$ is the label of the account $v_i$.

According to the above, each account $v_i$ is assigned an initial feature vector $x_i$, which is an essential input for most detection models. To capture the trading behavior patterns of the accounts, a set of concise and generic manual features is designed. This set consists of the following 17 features:

\begin{itemize}
  \item[$\bullet$] The total, average, maximum, minimum, and standard deviation of amounts for both received and sent transactions: $ 5\times 2 = 10$ types.
  \item[$\bullet$] The average minimum amount interval and time interval between consecutive received and sent transactions for an account: $2\times 2=4$ types.
  \item[$\bullet$] The lifecycle of an account: $1$ type.
  \item[$\bullet$] The number of transactions an account sends and receives: $1 \times 2 = 2$ types.
\end{itemize}
These features encompass various aspects of account activities, including fund flows, transaction frequency, distribution of transaction amounts, and time intervals, which effectively describe the transaction behavior patterns of accounts.
With these features, we model fraud detection in Ethereum as a node classification task on a graph. We establish a mapping from accounts representation to identity labels: $f(v,G) \mapsto y$. This mapping aims to identify potential fraud accounts by learning the accounts' behavior behaviors, thereby enhancing the security of the Ethereum ecosystem.

\subsection{Ethereum Interaction Hypergraph Modeling}

The homogeneous graph focuses on binary transactions between accounts, thereby neglecting the intricate multi-party interaction patterns that arise during contract invocations.
Consequently, we introduce a hypergraph modeling approach to capture the high-order interactions among accounts. Hypergraphs utilize hyperedges to simultaneously connect multiple nodes, offering a more flexible representation of complex transaction patterns within Ethereum. 
We express it as follows $G_H = (\mathcal V,\mathcal {E}_H, \boldsymbol H,\boldsymbol X, \mathcal Y)$, as shown in Fig.~\ref{fig1}(b).
$\mathcal{E}_H = \{e_{1},e_{2},\ldots ,e_{m}\}$ is a set of hyperedges, where each hyperedge $e_i$ contains multiple nodes, representing complex multi-party interactions which are associated through the same transaction hash. $\boldsymbol H \in \mathbb{R}^{n \times m}$ is the incidence matrix of the hypergraph, which represents the affiliation between nodes and hyperedges. The remaining components are consistent with those of a homogeneous graph.

In Ethereum, transaction hashes serve as unique identifiers for transactions, recording and tracking all transactional activities involved. Specifically, when accounts engage in direct transactions, hyperedges function similarly to regular edges by connecting two accounts.
However, when a transaction involves the smart contract call, a multi-party transaction pattern often emerges under a single transaction hash, as illustrated in Fig.~\ref{fig1}(b). 
Consequently, the hypergraph based on transaction hashes not only reflects individual transactional events but also reveals more comprehensive multi-party interaction information. 

\section{Method}
\label{method}

\begin{figure*}[t]
  \centering
  \includegraphics[width=\textwidth]{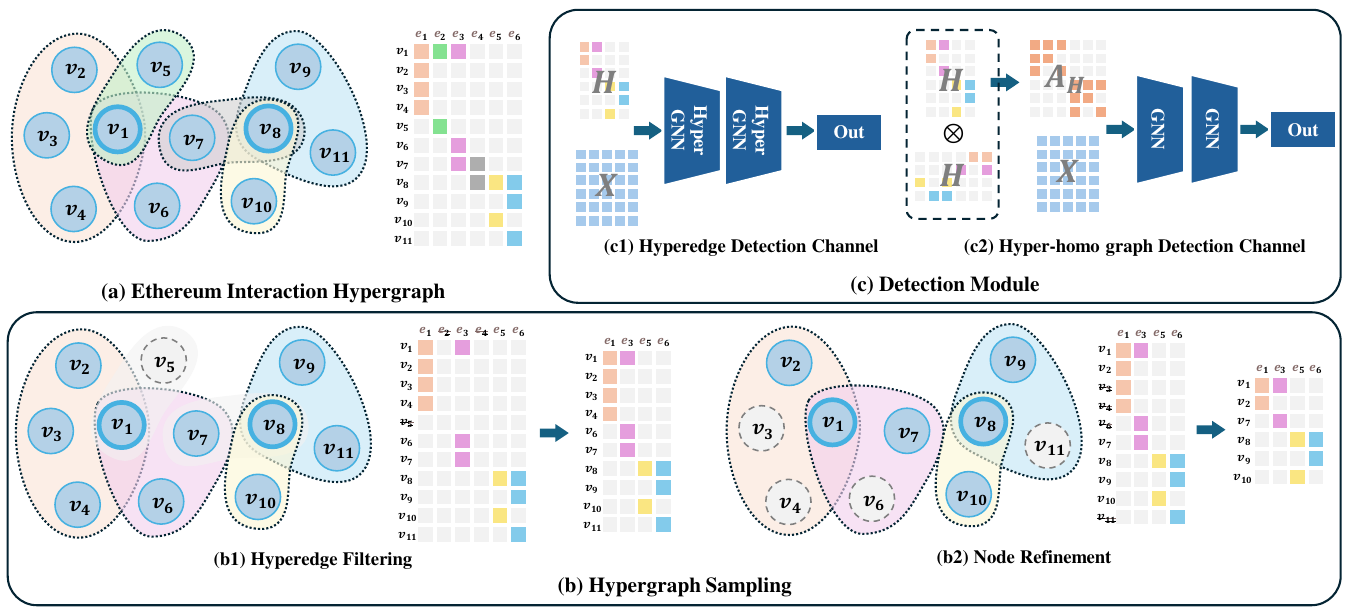}
  \caption{The framework of Ethereum interaction hypergraph sampling and dual-channel detection module.}
  \label{fig2}
\end{figure*}

In this section, we propose a two-step hypergraph sampling method and detection module, as illustrated in Fig.~\ref{fig2}. The proposed sampling method enables the generation of lightweight graphs, thereby facilitating efficient detection.

\subsection{Ethereum Interaction Hypergraph Sampling}
The two-step sampling method proposed consists of two phases: hyperedge filtering and node refinement, as illustrated in Fig.~\ref{fig2}(b). By employing this two-step sampling approach, we ensure the preservation of the structural integrity of the hypergraph while significantly reducing computational complexity. 

\subsubsection{Hyperedge Filtering}
In the context of a target node, there are multiple hyperedges within its neighborhood. To reduce computational complexity, we establish a   $\alpha$ to sample hyperedges around target nodes. Specifically, for a target node, its hyperedge set and the sampling strategy are defined as follows:

\begin{equation}
    \begin{gathered}
    \mathcal{E}_{H_i}=\{{e_i|v_i \in e_i}\}\\
    \hat{\mathcal{E}}_{H_i}= \begin{cases}\text {RandomSample}\left(\mathcal{E}_{H_i}, \alpha\right) & \text { if }\left|\mathcal{E}_{H_i}\right|> \alpha \\ \mathcal{E}_{H_i} & \text { if }\left|\mathcal{E}_{H_i}\right| \leq \alpha\end{cases}
    \end{gathered}
\end{equation}
where $\mathcal{E}_{H_i}$ denotes the set of all hyperedges containing node $v_i$, $\left|\mathcal{E}_{H_i}\right|$ denotes the number of hyperedges, $\text{RandomSample}(\mathcal{E}_{H_i}, \alpha)$ denotes the number of $\alpha$ hyperedges randomly sampled from $\mathcal{E}_{H_i}$, $\hat{\mathcal{E}}_{H_i}$ denotes the set of hyperedges obtained after the sampling. 
After the hyperedge filtering, each target node $v_i$ retains at most $\alpha$ hyperedges, as shown in Fig.~\ref{fig2}(b1). 
This sampling method not only reduces the amount of data but also preserves the higher-order interaction properties of the target node domain to some extent.

\subsubsection{Node Refinement}

Each hyperedge typically connects multiple nodes. To further reduce computational overhead, we introduce a threshold $\beta$ to sample the nodes within each hyperedge. Specifically, for each hyperedge $e_{i}$, node refinement is performed to sample the nodes $\mathcal{V}_{e_i}$ contained within it, as described by the following formula:

\begin{equation}
    \hat{\mathcal{V}}_{e_i}= \begin{cases}\left\{v_i\right\} \cup \text {RandomSample}\left(\mathcal{V}_{e_i} \backslash\left\{v_i\right\},\beta-1\right) & \text { if }\left|\mathcal{V}_{e_i}\right|>\beta \\ \mathcal{V}_{e_i} & \text { if }\left|\mathcal{V}_{e_i}\right| \leq \beta\end{cases}
\end{equation}
where $\left| \mathcal{V}_{e_i} \right|$ denotes the number of nodes contained in the hyperedge $e_i$, $\hat{\mathcal{V}}_{e_i}$ is the set of sampled nodes in the hyperedge, and $\text {RandomSample}\left(\mathcal{V}_{e_i} \backslash\left \{v_i\right\},\beta-1\right)$ denotes the random sampling of $\beta -1$ nodes other than the target node $v_i$ from the hyperedge $e_i$. 
After the inner node sampling stage, the number of nodes within each hyperedge is limited to $\beta$ as shown in Fig.~\ref{fig2}(b2), thus reducing the hypergraph scale. At the same time, this approach mitigates the issue of large discrepancies in the number of nodes across different hyperedges, ensuring more efficient learning in the subsequent detection phase.

\subsection{Dual-channel Detection Module}

After the two-step sampling, we construct a lightweight Ethereum interaction hypergraph. Subsequently, we design an efficient dual-channel detection module to be compatible with existing GNN- and HGNN-based detection algorithms, as shown in Fig.~\ref{fig2}(c). Both channels leverage the powerful modeling capacity of hypergraphs to further learn the latent interaction patterns of accounts and identify anomalous behaviors.

\subsubsection{Hypergraph Detection Channel}

The hyperedges, which serve as the medium that connects the accounts, are a pivotal component of the hypergraph message-passing process. By leveraging the hypergraph message-passing mechanism, it is possible to effectively ascertain the behavioral characteristics of the accounts within the trading network.
The hypergraph message passing first aggregates information from multiple nodes into their hyperedges to capture rich contextual information. Then, each account aggregates the hyperedge features from which it belongs. Finally, we obtain the final account representation. This mechanism effectively captures the local characteristics of accounts and can indirectly encode the high-level features through messages passing across hyperedges.

Utilizing the aforementioned hypergraph message passing mechanism, we are compatible with existing HGNNs, enabling the learning of account representations through the following formula:
\begin{equation}
    \boldsymbol{Z}_{\mathrm{hyper}}=\operatorname{GNN}_{\text {hyper }}(\boldsymbol{X}, \boldsymbol{H})
\end{equation}
where $\boldsymbol{Z}_{\mathrm{hyper}}$ denotes the account representation obtained from the hypergraph channel, and $\operatorname{GNN}_{\text {hyper}}$ denotes the existing hypergraph neural network model.

\subsubsection{Hyper-homo Graph Detection Channel}

Considering the high complexity of HGNNs and the fact that their applications are not yet mature enough compared to GNNs detection methods, we design a method to transform hypergraphs into hyper-homo graphs with the incidence matrix $\boldsymbol{H}$. 
The resulting hyper-homo graph can be compatible with existing homogeneous graph-based detection methods.
The specific process is delineated below:

\begin{equation}
    \begin{gathered}
\boldsymbol{A}_H=\boldsymbol{H} \cdot \boldsymbol{H}^T - \operatorname{Diag}\left(\boldsymbol{H} \cdot \boldsymbol{H}^T\right) \\
\boldsymbol{Z}_{\text {homo }}=\operatorname{GNN}_{\text {homo }}\left(\boldsymbol{X}, \boldsymbol{A}_H\right)
\end{gathered}
\end{equation}
where $\boldsymbol{Z}_{\text {homo}}$ denotes the account representation under the hyper-homo graph, $\operatorname{GNN}_{\text {homo}}$ denotes the GNN used to learn the network representation, and $\boldsymbol{A}_H$ is the hyper-homo adjacency matrix without self-loops, which preserves the higher-order association relations in the hypergraph.
Besides, it can be compatible with the already mature GNN algorithm to realize efficient fraud detection.

\subsection{Model Training and Optimization}
 
Following the implementation of two-channel detection, we obtain the account representations which are rich in multi-party interaction information. We regard the Ponzi detection as the node classification mission and use an MLP as a classifier to map the learned account representations to the classification space, which is formulated as follows:
\begin{equation}
\boldsymbol{P}=\operatorname{softmax}\left(\boldsymbol{W}\boldsymbol{Z}+\boldsymbol{b}\right)
\end{equation}
where $\boldsymbol{Z}\in \{\boldsymbol{Z}_\text{hyper}, \boldsymbol{Z}_\text{homo}\}$ are the account representations of the inputs of the different channels, $\boldsymbol{W}$ and $\boldsymbol{b}$ denote the weights and bias matrices, and the softmax function converts the outputs into probability distributions.
For the optimization objective, we use the cross-entropy loss function. The function effectively measures the difference between the predicted probability distribution and the true labels. The specific formula is as follows:

\begin{equation}
    \mathcal{L}=-\frac{1}{N} \sum_{i=1}^N \sum_{c=1}^C y_{i, c} \cdot \log \left(p_{i, c}\right)
\end{equation}
where $N$ is the number of samples, $C$ is the number of categories, $y_{i,c}$ is the true label of sample $v_i$, and $p_{i,c}$ is the predicted probability that the model predicts sample $v_i$ to belong to category $c$.

\section{Experiment}
\label{experiments}


\subsection{Dataset}
\begin{table}[t]
    \caption{Statistics of accounts, edges and hyperedges in different graphs.}
    \centering
    \renewcommand\arraystretch{1.2}      

    \resizebox{\textwidth}{!}{ 
    \begin{tabular}{c|cc|cccc} 
\hline\hline
\multirow{2}{*}{Graph} & \multicolumn{2}{c|}{Homogeneous Graph} & \multicolumn{4}{c}{Hypergarph}               \\
                       & Ori       & Samp                       & Ori        & Samp-S1 & Samp-S2 & Hyper-homo  \\ 
\hline
Node                   & 3,538,238 & 42,302                     & 3,538,238  & 76,325  & 47,887  & 47,887      \\
Edge/Hyperedge         & 5,027,678 & 75,665                     & 14,729,671 & 72,347  & 72,347  & 463,861     \\
\hline\hline
\end{tabular}
    }
    \label{tab: dataset}
\end{table}
We obtained 197 accounts labeled as Ponzi schemes and 1,406 normal accounts from \textit{Etherscan}\footnote[1]{\url{https://cn.etherscan.com}}. Based on these target nodes, we extracted their first- and second-order transaction and trace records during their most active years from \textit{Google BigQuery}\footnote[2]{\url{https://cloud.google.com/bigquery}}. 
We remove duplicate transactions and end up with 5,027,678 edges in the homogeneous graph.
For hypergraph, the original transaction hash forms 14,729,671 hyperedges. The statistical characteristics of the original and sampled dataset are summarized in Table~\ref{tab: dataset}.



\subsection{Experimental Setup}
For different graph modeling approaches, we evaluate the corresponding graph detection methods. For the homogeneous graph, we select three models including GIN, GCN, and SAGE. For hypergraph models, we select three models including HGNN, HNHN, and HyperGCN. We establish the model-related hyperparameters space, incorporating Boolean options \{\textit{True}, \textit{False}\} for batch normalization, \{16, 32, 64\} for hidden dimensions, and \{0.1, 0.05, 0.01, 0.005, 0.001\} for learning rate. The number of layers for GIN is fixed at 5, and for other models is fixed at 2. The dropout rate for all models is fixed at 0.5. The optimizer uses Adam for gradient descent, and the weight decay is fixed at 0.0005.  
For the hypergraph, the sampling threshold $\alpha$ is set to 100 in the hyperedge filtering step, and the sampling threshold $\beta$ in the node refinement step is selected from \{5, 6, 7, 8, 9\}.
For the homogeneous graph, to ensure the data alignment between the two graphs, we randomly sample 70 each of both first-order and second-order of the target. 
The grid search method is employed to obtain the best model parameters. 
All experiments are repeated five times, and the average performance value is presented to enhance the reliability and reproducibility of the results.
We divide the labeled data into 60\% for model training, 20\% for hyperparameter tuning and model selection, and the remaining 20\% for final performance evaluation.

To comprehensively evaluate the model performance, a variety of evaluation metrics are employed. The Binary metrics, including Precision, Recall, and F1-score, are utilized to evaluate the detection ability of positive samples. 
Besides, the F1-score under the Macro metrics and Area Under the Curve (AUC) are introduced to evaluate the comprehensive detection ability for the data imbalance. 
These metrics offer a multifaceted perspective on performance, thereby ensuring the scientific validity and reliability of the experimental results.

\begin{table}[t]
\renewcommand\arraystretch{1.2}      
\centering
\arrayrulecolor{black}
\caption{The results of Ponzi scams detection under different graphs for precision, recall, binary-f1, macro-f1, and AUC. The best results are bolded and the second best results are underlined.}
\resizebox{\textwidth}{!}{ 
\begin{tabular}{c|c|c|c|c|c|c|c} 
\hline\hline
Graph                                                                       & Methods   & Precision             & Recall                & Binary F1                    & Macro F1              & AUC                   & Rank                   \\ 
\hline
\multirow{3}{*}{\begin{tabular}[c]{@{}c@{}}Homogeneous\\Graph\end{tabular}} 
& GCN       & \uline{86.11 ± 2.03}  & 41.03 ± 1.62          & 55.54 ± 1.33          & 75.58 ± 0.70          & 70.05 ± 0.76          & \multirow{3}{*}{4.47}  \\
   & GIN       & 85.68 ± 4.01          & 48.72 ± 5.13          & 61.95 ± 4.42          & 78.98 ± 2.37          & 73.79 ± 2.53          &                        \\
  & SAGE & \textbf{88.10 ± 3.90} & 51.79 ± 2.99          & \textbf{65.12 ± 2.30} & \textbf{80.70 ± 1.24} & 75.40 ± 1.41          &                        \\ 
\hline
\multirow{3}{*}{Hypergraph}                                                 
& HGNN      & 67.57 ± 5.37          & 46.15 ± 2.81          & 54.71 ± 2.44          & 74.77 ± 1.36          & 71.52 ± 1.32          & \multirow{3}{*}{7.30}   \\
 & HNHN      & 67.05 ± 5.05          & \textbf{57.44 ± 4.76} & 61.86 ± 4.82          & 78.51 ± 2.70          & \uline{76.77 ± 2.63}  &                        \\
 & HyperGCN  & 73.76 ± 2.95          & 46.67 ± 10.93         & 56.48 ± 8.96          & 75.89 ± 4.72          & 72.20 ± 5.24          &                        \\ 
\hline
\multirow{3}{*}{\begin{tabular}[c]{@{}c@{}}Hyper-homo\\Graph\end{tabular}}  
& GCN       & 71.64 ± 5.50          & 49.74 ± 3.48          & 58.43 ± 1.15          & 76.81 ± 0.64          & 73.45 ± 1.26          & \multirow{3}{*}{4.07}  \\
  & GIN       & 79.83 ± 4.74          & 51.79 ± 1.92          & 62.75 ± 2.05          & 79.30 ± 1.16          & 74.98 ± 0.96          &                        \\
  & SAGE & 74.34 ± 3.27          & \uline{56.92 ± 6.36}  & \uline{64.38 ± 5.16}  & \uline{80.06 ± 2.82}  & \textbf{77.11 ± 3.24} &                        \\
\hline\hline
\end{tabular}
}
\label{tab: result}
\end{table}

\subsection{Evaluation of Hypergraph Modeling}
To verify the effectiveness of the hypergraph modeling proposed in this paper, we compare the existing graph-based detection methods. Incorporating the hyper-homo graphs proposed in this study, we ultimately construct three graph structures, including homogeneous graphs, hypergraphs, and hyper-homo graphs. We conduct a comparative analysis of the performance of fraud detection models based on these graph structures. As presented in Table \ref{tab: result}, a statistical evaluation of the comprehensive rankings of the three graph structures reveals that the detection performance based on hyper-homo graphs significantly outperforms that of homogeneous graphs and hypergraphs.

Specifically, for GCN and GIN, the hyper-homo graph outperforms the homogeneous graph across all evaluation metrics except precision. Meanwhile, although the SAGE model does not achieve state-of-the-art performance, it still ranks among the top two. This finding suggests that converting hypergraphs into hyper-homo graphs preserves the multi-party interaction patterns inherent in hypergraphs, thereby raising the upper bound of detection performance. Identifying a greater number of Ponzi schemes is critical for reducing economic losses. Thus, we consider recall a pivotal metric, as it reflects the proportion of detected Ponzi schemes. The result reports that the recall of the hyper-homo graph consistently surpasses that of the homogeneous graph. This indicates that hypergraph modeling based on transaction hashes effectively captures the behavioral patterns of fraudulent accounts, enhancing detection capabilities. Coupled with the strong performance in the F1-score, we deem the relatively lower precision acceptable. Furthermore, improvements in Macro F1 and AUC metrics substantiate that the hyper-homo graph structure mitigates the classification challenges posed by imbalanced data. In summary, the superior performance of the hyper-homo graph robustly validates the effectiveness of the transaction hash-based hypergraph modeling approach proposed in this study, underscoring the critical role of transaction hash information in fraud detection tasks.

Hypergraphs generally underperform homogeneous graphs across most metrics, exhibiting a relatively larger standard deviation. However, the robust performance of hyper-homo graphs demonstrates the effectiveness of the transaction hash-based hypergraph construction approach. Consequently, we attribute this phenomenon to the complexity of hypergraph models and their relative immaturity in the context of Ethereum detection applications. This observation motivates us to design more efficient hypergraph neural networks tailored to the specific characteristics of the Ethereum network.

\subsection{Computational Efficiency Analysis}

\begin{figure}
	\centering
	\includegraphics[width=0.8\textwidth]{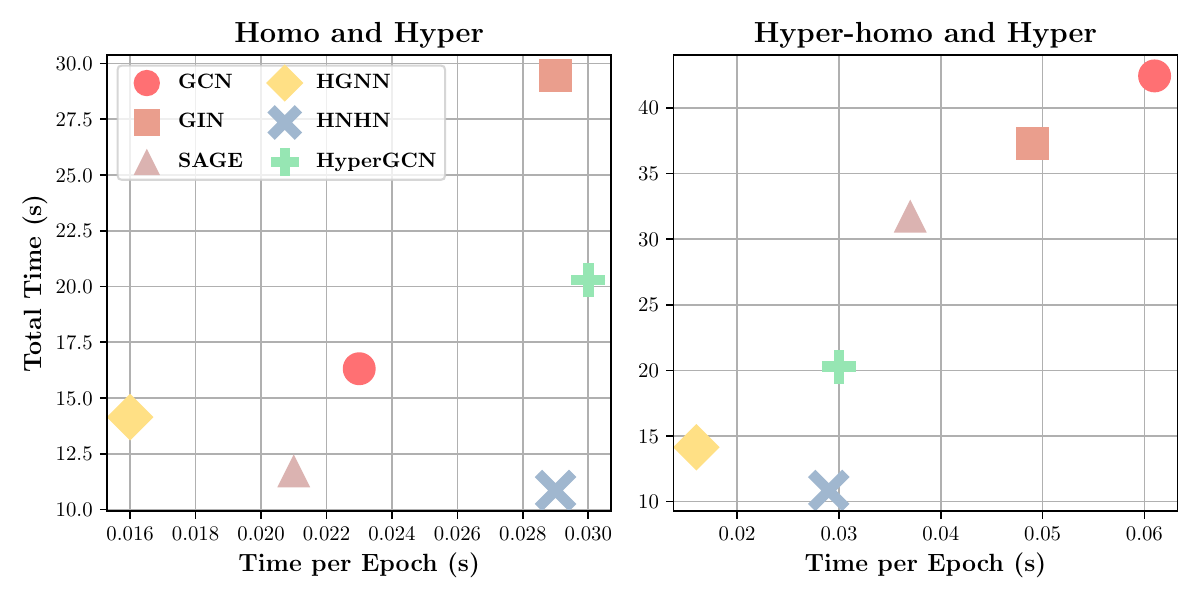}
	\caption{A comparison of the computational costs associated with various graphs.}
	\label{fig: time}
\end{figure}
As illustrated in Fig.~\ref{fig: time}, we analyze the computational costs associated with various graphs. The \textit{x}-axis denotes the training time of each epoch, while the \textit{y}-axis represents the total training time.  The left compares the performance of homogeneous graphs and hypergraphs, highlighting that the hypergraph model operates at a higher speed despite its more intricate message passing. Specifically, HGNN exhibits the shortest per epoch time, while HNHN demonstrates the least total training time, underscoring the efficacy of hypergraph models in application.  The right compares hypergraphs and hyper-homo graphs, revealing that the cost of hyper-homo graphs exceeds that of the hypergraph, which is attributed to the fact that the number of edges in hyper-homo graphs increases by approximately sevenfold.  
Despite the limitations of current hypergraph models in terms of detection performance, their message-passing architecture demonstrates significant practical value. We believe that the development of more efficient hypergraph models is a crucial direction for enhancing the performance and balancing the computational cost of Ethereum fraud detection.

\section{Conclusion} \label{conclusion}

This study focuses on transaction hash information to propose a hypergraph-based detection approach, called HyperDet, which could capture multi-party interaction patterns. 
We propose a two-step sampling method to reduce computational cost. 
We design a dual-channel detection framework to transform the hypergraph into the hyper-homo graph, which ensures compatibility with existing detection models. 
The experimental results demonstrate that the hypergraph works. However, HGNNs performs worse than GNNs, so we must develop an efficient HGNNs for Ethereum to improve detection.

\subsubsection*{Acknowledgments.} 
This work was supported in part by China Post-Doctoral Science Foundation under Grant 2024M762912, in part by the Post-Doctoral Science Preferential Funding of Zhejiang Province of China under Grant ZJ2024060, in part by the Key Research and Development Program of Zhejiang under Grants 2022C01018 and 2024C01025, in part by the National Natural Science Foundation of China under Grant U21B2001, and in part by the Key Technology Research and Development Program of Hangzhou under Grant 2024SZD1A23.

\bibliographystyle{splncs04_}
\bibliography{sample-base}

\end{document}